\documentclass[runningheads]{llncs}
\usepackage[T1]{fontenc}
\usepackage{xcolor}

\usepackage[square,comma,numbers,sort&compress,sectionbib]{natbib}
\usepackage{makecell}
\usepackage{adjustbox}
\usepackage{booktabs}
\usepackage{hyperref}
\usepackage{graphicx}
\usepackage{comment}
\usepackage{booktabs}
\usepackage{amsmath}
\usepackage{tabularx}
\usepackage{subcaption}    
\usepackage{pifont}
\usepackage[most]{tcolorbox}
\usepackage{rotating} 
\begin{document}
\title{Sim4IA-Bench: A User Simulation Benchmark Suite for Next Query and Utterance Prediction}
\titlerunning{Sim4IA-Bench: A User Simulation Benchmark Suite}

%
\author{
Andreas Konstantin Kruff\inst{1} \orcidID{0009-0002-8350-154X} 
\and
Christin Katharina Kreutz\inst{2} \orcidID{0000-0002-5075-7699} 
\and
Timo Breuer\inst{1} \orcidID{0000-0002-1765-2449} 
\and
Philipp Schaer\inst{1} \orcidID{0000-0002-8817-4632} 
\and
Krisztian Balog\inst{3} \orcidID{0000-0003-2762-721X}
}
\authorrunning{Kruff et al.}

\institute{TH Köln - University of Applied Sciences, Germany
\email{\{andreas.kruff,timo.breuer,philipp.schaer\}@th-koeln.de} \\ \and
TH Mittelhessen - University of Applied Sciences, Germany\\ \email{ckreutz@acm.org}
\\ \and
Stavanger University, Norway\\
\email{krisztian.balog@uis.no}}
\maketitle              
\begin{abstract}%
Validating user simulation is a difficult task due to the lack of established measures and benchmarks, which makes it challenging to assess whether a simulator accurately reflects real user behavior. As part of the Sim4IA Micro-Shared Task at the Sim4IA Workshop, SIGIR 2025, we present Sim4IA-Bench, a simulation benchmark suit for the prediction of the next queries and utterances, the first of its kind in the IR community. Our dataset as part of the suite comprises 160 real-world search sessions from the CORE search engine. For 70 of these sessions, up to 62 simulator runs are available, divided into Task A and Task B, in which different approaches predicted users’ next search queries or utterances. Sim4IA-Bench provides a basis for evaluating and comparing user simulation approaches and for developing new measures of simulator validity. Although modest in size, the suite represents the first publicly available benchmark that links real search sessions with simulated next-query predictions. In addition to serving as a testbed for next query prediction, it also enables exploratory studies on query reformulation behavior, intent drift, and interaction-aware retrieval evaluation.
We also introduce a new measure for evaluating next-query predictions in this task.
By making the suite publicly available, we aim to promote reproducible research and stimulate further work on realistic and explainable user simulation for information access: \url{https://github.com/irgroup/Sim4IA-Bench}.

\keywords{User simulation  \and Evaluation \and Next Query Prediction}
\end{abstract}
\section{Introduction and Motivation}

In recent years, user simulation has gained increasing attention within IR, as it provides a scalable and controllable method to study user behavior without large-scale user studies~\cite{INR-098,10.1145/3641525.3663619,DBLP:conf/sigir/OwoichoSA0C23}. The advent of large language models (LLMs) dramatically lowered the barrier to entry, making it easier than ever to create simulators capable of generating human-like search queries and conversational utterances~\cite{DBLP:conf/sigir-ap/Azzopardi00KM0P24,10.1145/3650041,10.1145/3726302.3730193}. However, this rapid development outpaces our ability to verify their performance. Consequently, there is little shared understanding of what constitutes a good simulator or how its performance should be evaluated~\cite{INR-098}. 

This challenge arises from a fundamental gap in evaluation methodology. The validation of user simulators is an open problem that requires two key components: (1) benchmark datasets that directly link real user interaction logs to simulated outputs, and (2) robust measures to quantify the similarity between simulated and real user behavior. Currently, there is a critical shortage of public resources dedicated to this task. Without a common ground for comparison, it is impossible to assess whether a new simulator is a true advancement or to understand the strengths and weaknesses of different simulation approaches.

To bridge this critical gap, this paper introduces Sim4IA-Bench, the first public benchmark resource specifically designed to evaluate user simulators. As the primary contribution of this work, we introduce a dataset derived from a recent user simulation initiative, the Sim4IA~\cite{DBLP:conf/sigir/SchaerKB0K25} Micro-Shared Task, tackling interactive IR simulation (Task A) and conversational session simulation (Task~B). In addition to submissions from participating approaches, we complement this dataset with a proposed set of string-based and system-based similarity measures, offering a crucial starting point for the community to assess simulator quality. 
Importantly, our goal is not to measure a simulator's success in a downstream retrieval task, but to directly address the more fundamental question of how well it reproduces authentic user behavior.

Beyond the methodological gap, a significant practical barrier has also hindered the wider adoption of user simulation: the substantial infrastructure and engineering effort required to build a simulator from scratch. To address this challenge, Sim4IA-Bench provides a comprehensive suite of practical resources designed to dramatically lower this barrier to entry. At the core, it includes a simulation toolkit that serves as a starting kit with baseline implementations, data loaders, and evaluation scripts. Sim4IA-Bench contains a rich collection of artifacts, such as prepared session logs, participants' run files, and comprehensive documentation detailing data formats and evaluation protocols. By packaging these components together, we shift the focus from foundational engineering and enable researchers to concentrate on the core scientific challenges of simulator design.
Sim4IA-Bench\footnote{GitHub repository of Sim4IA-Bench: \url{https://github.com/irgroup/Sim4IA-Bench}} is released under the MIT license, enabling both academic and industry researchers to access and use the resource.

\subsubsection{The Sim4IA-Bench Suite}

In addition to two session datasets for typical IR and conversational search from the academic domain, Sim4IA-Bench provides a comprehensive set of artifacts to support experimentation and evaluation:

\begin{itemize}
    \item Prepared session logs, including training and test sets.
    \item Submission run files (62) and corresponding lab notes from the three teams participating in the Sim4IA Micro-Shared Task.
    \item Benchmarking code for evaluating next-query prediction.
    \item A simulation toolkit, including Dockerized adaptations of SimIIR 3~\cite{DBLP:conf/sigir-ap/Azzopardi00KM0P24}.
    \item Tutorials and detailed documentation with setup instructions and example workflows.
\end{itemize}

Furthermore, this work is intended to guide the development of future community-wide evaluation initiatives. The methodology, dataset structure, and experiences gained from organizing this shared task provide a valuable blueprint for establishing larger-scale, standardized evaluation campaigns at TREC or CLEF, and Sim4IA-Bench will be maintained as part of the User Simulation subtask (Task~3) in LongEval@CLEF'26~\cite{longevalecir2026}.

\section{Related Work}

This section covers current directions for the validation of user simulators before datasets for next query and utterance prediction are presented.

\subsection{Validating User Simulators}

There is a current trend towards relying on simulation-based evaluation, especially through usage of LLMs~\cite{DBLP:conf/sigir/OwoichoSA0C23,10.1145/3708985,Wang:2023:EMNLP}. 
However, critical shortcomings can arise such as LLMs showcasing behavior that is unrealistic for humans~\cite{Davidson:2023:arXiv,Wang:2024:WWW} or a lack of natural variation that is usually found in human interactions~\cite{Terragni:2023:arXiv,Yoon:2024:NAACL}.

While simulators need to be validated against real human interactions~\citep{INR-098}, the specific requirements for simulators differ depending on what they are going to be used for, i.e. training vs. evaluation~\citep{Bernard:2024:ICTIR}. 
In general, such a comparison against human interactions may be performed at a distributional level, for example, by comparing (i) query characteristics (length, terms)~\citep{Azzopardi:2007:SIGIR}, similarity~\citep{DBLP:conf/ecir/KieselGMHS24}, or retrieval performance and shared task utility~\citep{DBLP:conf/ecir/BreuerFS22} for traditional search or (ii) the distribution of dialogue acts or success rate for conversational agents~\citep{Zhang:2020:KDD}.
Other approaches for validation include labeling specific instances according to different dimensions, like naturalness, usefulness, grammar for conversational utterances~\citep{Wang:2023:EMNLP,Sekulic:2022:WSDM,Zhang:2022:SIGIR} or comparing entire conversations (human vs. simulated) in a side-by-side manner~\citep{Zhang:2020:KDD,Wang:2023:EMNLP}.
Another method is the evaluation based on testers where testers are sets of IR systems over which a specific performance pattern can be expected that a simulator is trying to reproduce~\citep{DBLP:conf/sigir/LabhishettyZ21,DBLP:conf/ecir/LabhishettyZ22}.

While there are few resources dedicated to validating simulators, with Sim4IA-Bench we provide exactly this to bridge this gap.

\begin{table}[t]
\caption{Comparison of Interactive IR and Conversational Search datasets for next-query or next-utterance prediction.
\textbf{Size} refers to the number of sessions or conversations.
\textbf{T} informs whether the dataset covers traditional IR, \textbf{C} informs whether the dataset covers conversations.
\textbf{A} refers to additional public assets like system runs from a shared task.
\textbf{Domain} indicates the topic of the dataset.}
\label{tab:related}
\begin{adjustbox}{width=\textwidth}
\begin{tabularx}{\textwidth}{
  l 
  l
  *{3}{>{\centering\arraybackslash}X} 
  p{4.25cm}
}
\toprule
Dataset &
Size &
T&
C &
A &
Domain \\
\midrule
AOL \cite{DBLP:conf/infoscale/PassCT06} & 283,207 (AOL17) & \ding{51} & \ding{55} & \ding{55} & Web search \\
SUSS \cite{DBLP:conf/ercimdl/MayrK17} & 484,449 & \ding{51} & \ding{55} & \ding{55} & Academic search \\
Yandex \cite{Serdyukov2014} & 797,867 & \ding{51} & \ding{55} & \ding{55} & Web search \\
TREC Session \cite{TREC2014} & 1564 & \ding{51} & \ding{55} & \ding{51} & Web search \\
ConvAI \cite{10.1007/978-3-319-94042-7_3} & 4750 & \ding{55} & \ding{51} & \ding{51} &Human Chatting  \\
TianGong-ST \cite{DBLP:conf/cikm/ChenMLZM19} & 147,155 & \ding{51} & \ding{55} & \ding{51} & Web search \\
ConvAI2 \cite{dinan2019secondconversationalintelligencechallenge} & 4406 & \ding{55} & \ding{51} & \ding{51} & Human Chatting \\
TripClick \cite{Rekabsaz2021} & 1,602,648 & \ding{51} & \ding{55} & \ding{51} & Health  \\
ConvAI3 (ClariQ) \cite{aliannejadi-etal-2021-building} & 1,596,757 & \ding{55} & \ding{51} & \ding{51} & Human Chatting \\
Baidu-ULTR \cite{zou2022} & 1.2 bil & \ding{51} & \ding{55} & \ding{51} &  Web search \\
Webis-FUQ-24 \cite{DBLP:conf/ecir/KieselGMHS24} & 18,980 & \ding{51} & \ding{55} & \ding{55} & Web search, arguments, exhibitions, product search\\
Persona-Chat \cite{zhang-etal-2018-personalizing} & 10,907 & \ding{55} & \ding{51} & \ding{55} & Human chatting\\
Webis-CQR-2 \cite{DBLP:conf/desires/KieselCB0H21} & 284 & \ding{55} & \ding{51} & \ding{55} & Arguments, books, news, trips\\
SoguoQ \cite{Song2021} & 14,075,717 & \ding{55} & \ding{51} & \ding{51} & Web search\\
TREC CAsT (2022) \cite{Owoicho2022TRECC2} & 50 & \ding{55} & \ding{51} & \ding{51} & Web search\\
LLM-REDIAL \cite{DBLP:conf/acl/LiangJWFXCY24} & 47,600 & \ding{55} & \ding{51} & \ding{55} & Movies, books, sports \\
WildChat \cite{Zhao2024WildChat1C} & 1,039,785 & \ding{55} & \ding{51}  & \ding{55} & Web search\\
LMSYS-CHAT-1M \cite{a2ae59163aba4853b175b99ae00c9990}& 1,000,000 & \ding{55} & \ding{51} & \ding{55} & Web search \\
\midrule
Ours & 160 &  \ding{51} & \ding{51} & \ding{51} & Academic search \\ 
\bottomrule
\end{tabularx}
\end{adjustbox}

\end{table}

\subsection{Datasets}

To the best of our knowledge, our introduced resources are the first to provide a common evaluation environment for both interactive IR as well as conversational search. To highlight the novelty of our resource, we surveyed existing log datasets and conversational resources containing session interaction logs. \autoref{tab:related} provides an overview of publicly available session datasets at the time of our study.

There are several large-scale datasets that primarily originate from the web search domain or rather small-scale domain-specific datasets from academic, health-related, and other fields. Most of these datasets serve as evaluation tool\-kits for different aspects of the user modeling in an interactive search setting with varying degrees of an explicit user simulation. For instance, some datasets allow a comprehensive evaluation of the different interactions in a simulated sessions, while others have a more specific focus like query suggestion, click modeling, or utterance prediction. Building on these datasets, prior work has examined related aspects such as query expansion and suggestion~\cite{DBLP:conf/sigir/Mitra15,DBLP:conf/sigir/ChenCCR18}, query and utterance prediction~\cite{wang-etal-2025-know,ivey2024realroboticassessingllms, Liu2020YouIM,10.1145/3578337.3605143}, or session modeling \cite{DBLP:conf/www/ChengRLRCLR21, DBLP:conf/cikm/ChenD0CCW22, DBLP:conf/aaai/YeLD0ZWC23, DBLP:conf/ercimdl/GuntherGH21}.

The additional assets (A) column in \autoref{tab:related} denotes whether a dataset has previously been employed in a shared task or is included in a benchmark that enables systematic evaluation.
Several of the listed datasets have served as complementary assets in this sense.
For example, the Yandex dataset was used in a challenge focused on personalizing search results based on user context and search history \cite{Serdyukov2014}.
The four TREC Session datasets were employed in a task designed to improve retrieval effectiveness through the use of historical queries, ranked result lists, and user interaction information \cite{TREC2014}.
Similarly, the Baidu-ULTR dataset was used in a task where participants developed feature-based re-ranking models that exploited behavioral and display features to better capture user preferences \cite{niu2023ultre2}.
The SogouQ dataset was used in a task addressing ambiguous queries and promoting ranking diversification to account for multiple possible user intents \cite{Song2011OverviewOT}.
TianGong-ST has been applied in a task focused on ranking documents for the final query of a session, taking into account the complete preceding session context \cite{chen2022ss}.
More recently, dialogue datasets such as WildChat \cite{Zhao2024WildChat1C} have been designed to evaluate large language models in realistic conversational settings \cite{DBLP:conf/iclr/LinDCRPD0025,DBLP:journals/corr/abs-2509-17442}.

Taken together, these datasets exemplify how complementary assets facilitate structured comparison, either by providing directly comparable shared task runs or by being integrated into established benchmarks.
Building on this, our resource is the first to provide complementary assets specifically for next query prediction in the context of user simulation, offering run files that enable reproducible system comparisons and systematic evaluation of measures under controlled experimental conditions.

In general, there is a trade-off between the desire to make as much user interaction data available for rigorous validation of user simulations and the requirements to keep users anonymous and respecting their privacy. Sim4IA-Bench enables validations of user simulations across multiple sessions in traditional IR and conversational search, while guaranteeing full user privacy based on rigorous anonymization measures.

\section{Dataset}

\begin{figure}[t]
    \centering
    \includegraphics[width=1\textwidth]{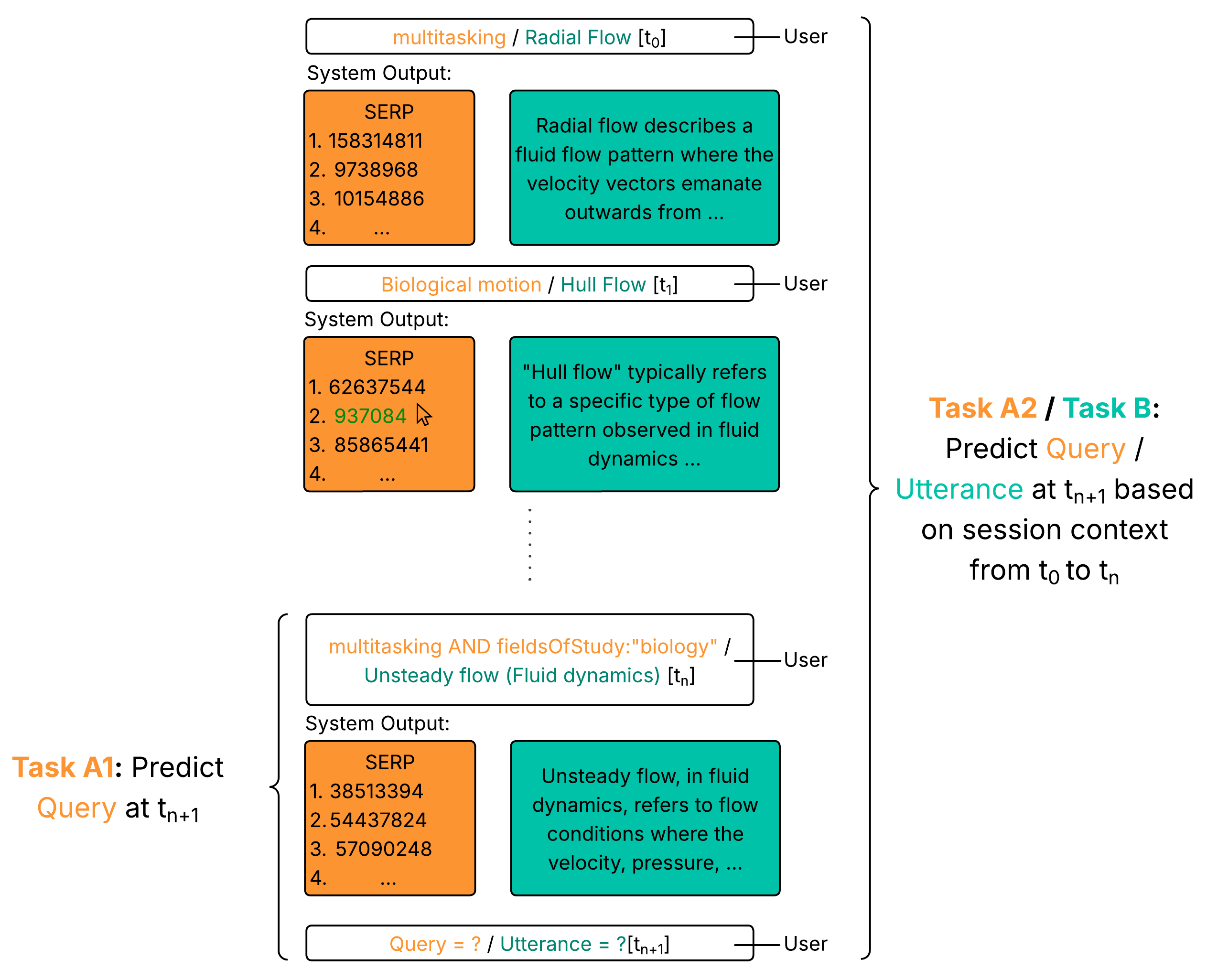}
    \caption{Overview of Task A (interactive IR simulation, in orange) and Task B (conversational session simulation, in teal). Document IDs highlighted in green represent documents a real user interacted with.}
    \label{fig:task_desc}
\end{figure}

Here we present the dataset at the core of Sim4IA-Bench: its structure, contents, and key characteristics, highlighting the aspects that make it suitable for evaluating and developing user simulators in interactive IR.

\subsection{Task Descriptions and Dataset Contents}

We provide session-based datasets for the Sim4IA Micro-Shared Task~\cite{DBLP:conf/sigir/SchaerKB0K25}, designed to support the evaluation and development of user simulators in interactive IR. The resources include two sets of data corresponding to the two task variants: Task A (interactive IR simulation) and Task B (conversational session simulation). Each task has a training set of 45 sessions and a test set of 35 sessions. The test sets differ from the training sets in that the final query of each session is withheld, allowing participants to evaluate simulator predictions without overfitting. In all tasks ten next queries or utterances are to be predicted.
For a concise overview of the two tasks and their respective workflows, see~\autoref{fig:task_desc}.

\subsubsection{Task A: Interactive IR Simulation} 
Each session in Task A contains queries, the corresponding retrieved SERPs, and the documents clicked by users. Clicks are categorized into three types: clicks on authors, clicks on the work itself, and clicks on the ``Download PDF'' option. Timestamps for queries and clicks are included, enabling simulators to account for temporal aspects of interactions and to model different types of user behavior. We composed two variants of Task A using the same data: Task A1 only considers the last query and corresponding SERP information, while in Task A2, the whole session was allowed to be used.
For Task A, the dataset contains an average session length of 5.20 queries (4.20 reformulations on average), with 1.49 clicks per query. 

\subsubsection{Task B: Conversational Session Simulation} 
Task B sessions contain only utterance-response pairs. Responses were generated using Google's Gemma 3 12B model served on Ollama. For each query, the model was provided with the top three retrieved documents, the preceding response, and all upcoming utterances. The prompt (see \autoref{prompt}) was designed to ensure that the model focused solely on the current utterance while steering towards the upcoming utterances, producing coherent and plausible conversational flows.
For Task B, the average session length is 4.85 queries (3.85 reformulations on average).  

\begin{figure}[t]
\begin{tcolorbox}

\textcolor{black}{Utterance:} \textcolor{purple}{What are the main symptoms of diabetes?}\\[1mm]
\textcolor{black}{Relevant Documents:}\\
\textcolor{black}{Title:} \textcolor{purple}{Diabetes Overview}\\
\textcolor{black}{Abstract:} \textcolor{purple}{Diabetes is a chronic condition characterized by high blood sugar levels. Common symptoms include frequent urination, excessive thirst, fatigue, and blurred vision.}\\
\textcolor{black}{Title:} \textcolor{purple}{Type 2 Diabetes Symptoms}\\
\textcolor{black}{Abstract:} \textcolor{purple}{Type 2 diabetes often develops gradually. Symptoms include increased hunger, frequent infections, and slow-healing sores.}\\[1mm]
\textcolor{black}{Previous Response:} \textcolor{purple}{Diabetes is a condition affecting the body's ability to regulate blood sugar.}\\
\textcolor{black}{Upcoming User Utterances:}\\
\textcolor{black}{Utterance:} \textcolor{purple}{How can diabetes be managed effectively?}\\
\textcolor{black}{Utterance:} \textcolor{purple}{What lifestyle changes are recommended for patients?}\\[1mm]
\textcolor{black}{Instruction:}\\
\textcolor{black}{- Answer the query using the relevant documents and the previous response.}\\
\textcolor{black}{- Act like a RAG system: provide relevant, informative, and context-aware responses tailored to the user’s query intent.}\\
\textcolor{black}{- Try to only answer the current utterance and explicitly do provide answers in the response, that might answer the provided upcoming utterances.}\\
\textcolor{black}{- Limit your answer to \textbf{no more than 150 words}.}\\
\textcolor{black}{- Focus on key points and avoid unnecessary repetition.}\\
\textcolor{black}{- Please do answer formally and don't use phrases like 'Okay, here's a response to ``Diabetes''' acting as a RAG system.}\\

\textcolor{black}{Answer:}

\end{tcolorbox}
\caption{Prompt template for generating the synthetic responses of the conversational system for Task B. Black parts show the unchanging structural and instructional parts, purple components depict the variables inserted depending on sessions and utterances.}
\label{prompt}
\end{figure}

\subsection{Session Extraction from CORE Logs}

Sessions were reconstructed from CORE \cite{knoth2023core} log files using a heuristic approach to group queries into sessions: starting from the last query in a potential session, queries occurring within a time window of -10 to +5 minutes were considered candidates for inclusion. Queries were added to a session if their cosine similarity, calculated using embeddings generated by the SentenceTransformer model all-MiniLM-L6-v2, with the current session queries, was greater than or equal to 0.1. All reconstructed sessions were independently manually reviewed by two people to ensure coherence and plausibility. Query sessions that contained fewer than three reformulations or did not originate from a plausible information need were excluded. 

\subsection{Usage Notes}

The datasets and artifacts are designed to be fully accessible and ready for use. Example scripts and tools facilitate loading, processing, and analysis of the sessions. For Task A, the combination of training and test sets allows evaluation of next-query predictions at multiple levels, from string similarity to semantic and SERP-based metrics. Task-specific features, such as click categories in Task A and semi-synthetic responses in Task B, enable more nuanced simulator evaluations and the exploration of interaction dynamics in both interactive and conversational settings.

\subsection{Submitted Runs}

For the Micro-Shared Task we obtained 62 runs over all tasks in total and three accompanying lab notes detailing information on these runs by groups from \texttt{CIR}~\cite{busch_2025_16909638}, \texttt{Webis}~\cite{gohsen_2025_16909542} and \texttt{THM}~\cite{dietzler_2025_17386068}.
In general, we distinguish between runs that were composed (semi-) manually, through an LLM, as well as rule-based ones. We further differentiate LLM-based runs as ones composed by prompting the LLM to behave as a persona alone, through a combination of tuning and prompting and others. 
\autoref{tab:run_types} provides the number of runs of different types from each group for the respective tasks. 

\begin{table}[t]
    \centering
        \caption{Distribution of submitted run types for tasks by groups from \textcolor{magenta}{\texttt{CIR}}, \textcolor{cyan}{\texttt{Webis}} and \textcolor{olive}{\texttt{THM}}.}
    \label{tab:run_types}
\begin{adjustbox}{width=\textwidth}
\begin{tabularx}{\textwidth}{
  l| 
  *{5}{>{\centering\arraybackslash}X}|r 
}    \toprule
         & (semi-) manual & persona & prompting \& tuning & other LLM & rule-based & $\sum$\\ 
        \midrule
        Task A1 & \textcolor{olive}{3}& \textcolor{olive}{6} &\textcolor{magenta}{2} + \textcolor{cyan}{3} = 5 & \textcolor{magenta}{3} + \textcolor{olive}{12} = 15  &\textcolor{magenta}{1} & 30\\
        Task A2 & \textcolor{cyan}{2} + \textcolor{olive}{3} = 5& \textcolor{olive}{6} &\textcolor{cyan}{3}& \textcolor{olive}{4} & & 18\\
        Task B & \textcolor{olive}{4} & \textcolor{olive}{6} &&\textcolor{olive}{4}&&14\\
        \bottomrule
    \end{tabularx}
    \end{adjustbox}

\end{table}

\section{Assessing Simulator Fidelity}

Evaluating the quality and validity of a simulator in query prediction tasks is challenging, and the choice of suitable measures remains an open question. In this work, our methodology shifts the focus from a simulator's effectiveness in retrieval tasks to its \emph{reproduction quality}---that is, how well it replicates the authentic user behavior observed in real interaction logs. To assess this, we employ a complementary set of string-based and system-based similarity measures. Each measure is designed to capture a different aspect of simulator fidelity by quantifying the degree to which simulated queries match actual user inputs. In addition to the following measures presented in this paper, a broader overview of suitable measures was conducted in the work of Kruff et al in an additional comprehensive study~\cite{Kruff:2026:ECIR}.

\subsection{Measures}

\textbf{Semantic Similarity.} Semantic similarity assesses how close the predicted queries are in meaning to the original user queries. We computed the average cosine similarity between sentence embeddings of the original query $q_i^{true}$ and the $Q$ candidate queries $q_{i,j}$ over $N$ sessions. We then bounded the similarity values from $[-1, 1]$ to $[0, 1]$ for better visual comparability with the other measures.

\[
\bar{S} = \frac{1}{|N|} \sum_{i=1}^{N} \left( \frac{1}{Q} \sum_{j=1}^{Q} \text{cosine}(q_i^{true}, q_{i,j}) \right)
\]

This measure ensures that candidates are not only syntactically similar but also semantically aligned with the user’s intent.

\textbf{Redundancy.} To evaluate novelty and diversity, we introduced a Redundancy measure, which calculates the Jaccard similarity between all $Q$ candidate queries or utterances within a session averaged across all $N$ sessions:

\[
\bar{R} = \frac{1}{|N|} \sum_{i=1}^{N} \left( \frac{2}{Q(Q-1)} \sum_{1 \leq j < q \leq Q} \text{Jaccard}(q_{i,j}, q_{i,k}) \right)
\]

Low redundancy indicates that a simulator produces multiple distinct candidates that remain semantically similar to the original query, whereas high redundancy signals minimal variation. Redundancy is not intended as a stand-alone metric; rather, it is defined for its role in the Rank-Diversity Score described below. It serves to penalize simulators that produce candidates with only minor variations, while giving simulators the opportunity to also predict queries exhibiting a degree of in-session topic drift.

\textbf{SERP Overlap.} To capture system-level effects, we measured the average overlap between the search engine results retrieved for the original query and the $Q$ candidates over the $N$ sessions:

\[
\bar{O} = \frac{1}{|N|} \sum_{i=1}^{N} \left( \frac{1}{Q} \sum_{j=1}^{Q} \text{overlap}(q_{i}^{\text{true}}, q_{i,j}) \right)
\]

Overlap denotes the fraction of shared documents in the top-10 ranked results by a fixed retrieval system. In our experiments, queries were executed using BM25. This metric assumes access to a search engine and corpus on which both original and candidate queries can be run, and is therefore dependent on the underlying retrieval model and index. The cutoff was set with reference to typical precision scores and the typical length of a webpage in the absence of pagination.

Accordingly, SERP overlap should be interpreted as a system-dependent indicator of how query reformulation affects downstream retrieval behavior, complementing string- and meaning-based similarity metrics rather than serving as a standalone effectiveness measure.

\textbf{Rank-Diversity Score.} While the above measures capture overall candidate quality, they do not consider the ordering of candidates. To address this, we designed an MMR-inspired Rank-Diversity Score (RDS) that combines rank-based evaluation with redundancy:

\[
\bar{RD} = \text{RDS}_{\cos\ge 0.7} \cdot (1 - \bar{R})
\]
\[
\text{RDS}_{\cos\ge 0.7} = \frac{1}{|N|} \sum_{i \in N} \left( \sum_{q \in Q_i} \frac{1}{\text{rank}_q^{\cos\ge 0.7}} \right)
\]

RDS is computed over all sessions $N$ and the list of candidates $Q_i$ generated for each session $i$. It rewards simulators that rank high-quality, diverse candidate queries or utterances at the top while penalizing poor ordering and runs submitting fewer than the required ten candidates, as this lowers their potential multiplier. Originally introduced by Carbonell and Goldstein~\cite{DBLP:conf/sigir/CarbonellG98}, MMR provides a conceptual foundation for our measure.

\subsection{Case Study: Analysis of Task A1}

\begin{figure}[t]
    \centering
    \includegraphics[width=1\textwidth]{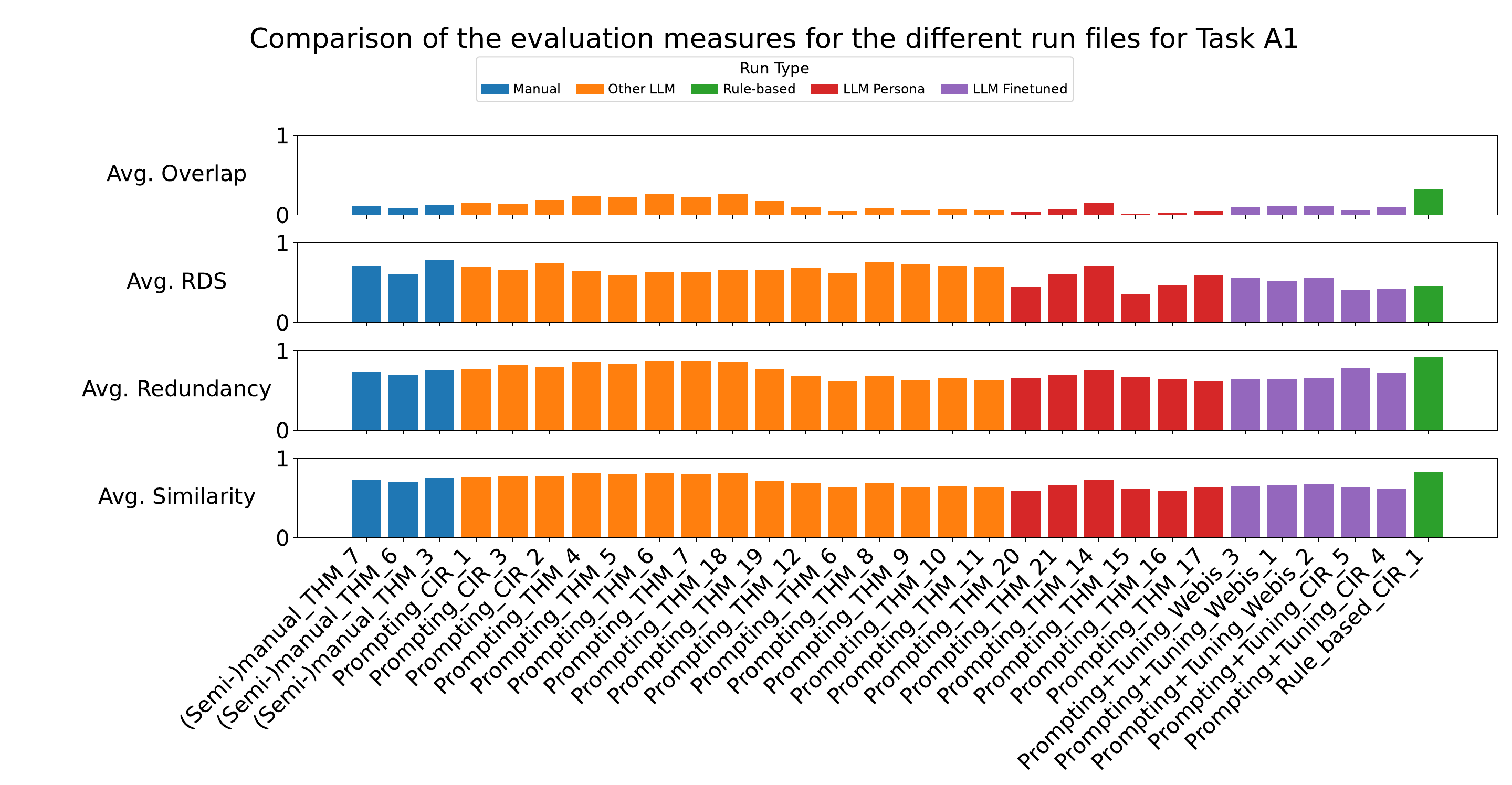}
    \caption{Exemplary run results for Task A1.}
    \label{fig:overall}
\end{figure}

Using the runs from Task A1 as an example, we illustrate how the applied measures provide complementary perspectives on simulator performance. Cosine similarity captures the semantic closeness between simulated and actual user queries, showing that manually created runs and large, out-of-the-box LLMs perform comparably well in reproducing the linguistic and conceptual intent of users. Persona-based and fine-tuned simulators, by contrast, show slightly lower alignment, while rule-based approaches reach high similarity scores due to their strong adherence to the original queries.

SERP overlap extends this perspective to the system level, reflecting whether different simulators lead to comparable retrieval outcomes. Here, the rule-based approach again performs strongest, as its low deviation from the original queries results in highly similar retrieval results. However, this measure also highlights that approaches generating more diverse queries, such as manual, persona-based, or fine-tuned runs, tend to diverge in retrieval outcomes, which suggests a trade-off between fidelity and behavioral variability.

Finally, the redundancy-based analysis, incorporated into the newly proposed measure, reveals a key limitation of several simulator types. Their tendency to produce highly similar candidate queries with little internal variation becomes particularly apparent for rule-based and some prompting-based approaches as well as for fine-tuned models that closely follow prior query patterns. In contrast, manual and large LLM runs benefit from this measure, as they generate more diverse candidate sets while maintaining overall semantic coherence.

Overall, while cosine similarity and SERP overlap are well suited for assessing how closely simulators mirror real user intent and retrieval outcomes, the redundancy-oriented measure complements them by exposing a lack of diversity across candidate queries. Together, these measures provide a multifaceted understanding of simulator behavior, revealing the strengths of faithful reproduction and the weaknesses in behavioral variability across approaches.

\subsection{Reuse Value of the Benchmark Artifacts}

Ultimately, Sim4IA-Bench establishes a foundation for developing future validation measures for user simulations. The corresponding process would follow standard practices used in developing IR evaluation metrics. For example, methods such as swap counting~\cite{DBLP:conf/sigir/VoorheesB02}, stability tests~\cite{DBLP:conf/sigir/BuckleyV00}, or bootstrapping~\cite{DBLP:conf/sigir/Sakai06}  could be applied to determine whether a new validation measure aligns with existing ones, including those introduced in this work, or opens up complementary evaluation perspectives. In this context, the focus of the validation would shift from ranking systems to ranking user simulators instead. 

\section{Reflections on the Shared Task}
\label{sec:reflections}

The considerable effort invested in preparing the study environment, including comprehensive setup instructions and video guides, proved successful. None of the participating teams required substantial support from the organizers. The smooth onboarding process was a key logistical success, ensuring that the setup can be easily reproduced in future studies.

The shared task and its resulting dataset provide a valuable first step toward systematically exploring user simulation in interactive search. However, the process also highlighted several key challenges and limitations. A significant finding, particularly for Task B (conversational session simulation), was that the simulated utterances remained largely ``query-like'' and lacked the natural verbosity typical of this setting. This points to a major area for future improvement in simulator design.
Furthermore, the semi-synthetic nature of the dataset highlighted both possibilities and constraints, offering opportunities for experimentation while revealing areas where realism could be improved.
Finally, the task underscored that selecting appropriate evaluation measures is still an open and underexplored challenge for the community.

\section{Conclusion and Outlook}

This work addresses the critical need for standardized evaluation of user simulators in IR.
We introduce Sim4IA-Bench, a comprehensive suite of resources derived from the Sim4IA~\cite{DBLP:conf/sigir/SchaerKB0K25} shared task, which includes datasets, task definitions, a baseline toolkit, tutorials, and documentation. This suite represents an initial yet significant step toward advancing the systematic study and evaluation of user simulators. By providing a structured, reusable framework, they enable reproducible experimentation and support methodological exploration in interactive IR. It also facilitates the testing of alternative simulation approaches and evaluation measures, inspiring the development of new tasks, datasets, or experimental designs.

This resource is not a static endpoint but a foundation for future community efforts. As an immediate next step, the methodology and toolkit will be maintained and expanded as part of the User Simulation subtask (Task 3) in LongEval@CLEF'26.
This transition from a micro-shared task to a full, recurring shared task will facilitate the collection of new datasets (starting with new data for Task A) and the evaluation of new measures, providing further insights into simulation-based evaluation of next-query prediction.

Looking ahead, the limitations identified in our reflections (Section \ref{sec:reflections}) define a clear research agenda.
Future work can build on this foundation to create richer, more realistic simulations that capture the full spectrum of user behaviors, especially for conversational search.
The resource provides a practical basis for testing and refining new user simulators, exploring alternative evaluation measures, and investigating how simulators generalize across different tasks and systems. As a framework, it enables reproducible experiments, highlights the trade-offs between realism, control, and evaluative rigor, and helps the community identify best practices. By making this resource accessible to the research community, we hope to encourage broader adoption, systematic benchmarking, and iterative improvement of simulation-based methods in interactive IR. Ultimately, the experiences gained from this initiative serve as a blueprint for our long-term goal: to establish a dedicated shared track at a major venue like TREC or CLEF, focused entirely on the broader research challenges of validating user simulators.

\begin{credits}
\subsubsection{\ackname}
This work is partially funded by Deutsche Forschungsgemeinschaft (DFG) under grant number 509543643 and within the funding programme FH-Personal (PLan CV, reference number 03FHP109) by
the German Federal Ministry of Education and Research (BMBF) and Joint Science Conference (GWK).

\subsubsection{\discintname}
The authors have no competing interests to declare that are relevant to the content of this article.

\end{credits}

\bibliographystyle{splncs04nat}
\bibliography{sim4ia.bib}

\end{document}